\newif\ifcameraready
\camerareadyfalse

\documentclass[a4paper,USenglish,cleveref,autoref]{lipics-v2021}

\ifcameraready
\else
  \hideLIPIcs
\fi
\nolinenumbers

\usepackage{amsmath,amssymb,amsfonts}
\usepackage{algorithm}
\usepackage{algpseudocode}
\usepackage{graphicx}
\usepackage{textcomp}
\usepackage{xcolor}
\definecolor{classicalStrategy}{HTML}{2166AC}
\definecolor{conservativeStrategy}{HTML}{B2182B}
\usepackage{marvosym}
\renewcommand{\mailsymbol}{\textcolor{lipicsGray}{\fontsize{9}{12}\selectfont\Letter}}
\renewcommand{\orcidsymbol}{\includegraphics[height=9pt]{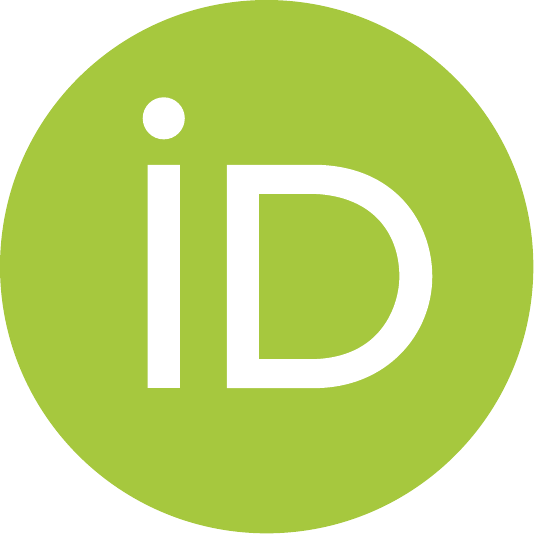}}
\usepackage{soul}
\usepackage{listings}
\usepackage{comment}
\usepackage{colortbl}         %
\usepackage{longtable}
\usepackage{arydshln}         %
\usepackage{booktabs}
\usepackage{tabularx}
\usepackage{array}
\usepackage{seqsplit}
\usepackage{adjustbox}
\usepackage{needspace}

\newcolumntype{L}[1]{>{\raggedright\arraybackslash}p{#1}}

\usepackage{tikz}
\usetikzlibrary{arrows.meta, chains, positioning, shapes.geometric}

\title{Inside Qubic's Selfish Mining Campaign on Monero: Evidence, Tactics, and Limits}
\titlerunning{Inside Qubic's Selfish Mining Campaign on Monero}

\author{Suhyeon Lee}{ShardLab, Seoul, Republic of Korea \and Hashed Open Research, Seoul, Republic of Korea}{orion-alpha@korea.ac.kr}{https://orcid.org/0000-0002-1318-6612}{}
\author{Hyeongyeong Kim}{Korea University, Seoul, Republic of Korea}{}{}{}
\authorrunning{S. Lee and H. Kim}
\Copyright{Suhyeon Lee and Hyeongyeong Kim}


\category{}
\ifcameraready
  \relatedversiondetails{Full Version}{https://arxiv.org/abs/2512.01437}
  \supplementdetails[linktext={GitHub repository}]{Datasets and Analysis Code}{https://github.com/shlee-lab/Qubic-selfish-mining-study}
\fi
\acknowledgements{We are grateful to the Monero community for sharing artifacts and empirical observations that complemented our measurements. We especially thank DataHoarder (\url{https://github.com/WeebDataHoarder}) and Sergei Chernykh (sech1) for providing access to supplemental datasets and answering technical questions that helped us validate and refine parts of our analysis. We thank the anonymous reviewers for their thoughtful feedback, which helped improve the paper.}

\EventEditors{Aggelos Kiayias and Maria Kyropoulou}
\EventNoEds{2}
\EventLongTitle{8th Conference on Advances in Financial Technologies (AFT 2026)}
\EventShortTitle{AFT 2026}
\EventAcronym{AFT}
\EventYear{2026}
\EventDate{October 6--9, 2026}
\EventLocation{London, United Kingdom}
\EventLogo{}
\SeriesVolume{395}
\ArticleNo{0}

\begin{document}

\hypersetup{pdfsubject={Accepted at the Conference on Advances in Financial Technologies (AFT 2026)}}

\maketitle

\begin{abstract}
In 2025, the blockchain network Qubic conducted a mining campaign on Monero, a privacy-focused cryptocurrency. It publicly presented the campaign as a ``51\% takeover'' and a demonstration of selfish mining. The episode provides a rare opportunity to test such claims in a privacy-preserving proof-of-work network, where pool attribution and private release decisions are difficult to observe. We combine Monero node measurements, Qubic pool observations, community-shared artifacts, and disclosed view keys to attribute blocks and identify candidate withholding periods. Rather than sustained majority control, we observe elevated orphaning and deeper reorganizations. Qubic did not follow a single optimized selfish mining strategy, instead varying between standard selfish mining and more defensive block-release decisions. Across the campaign, it gained no reward advantage over honest mining. Community monitoring and Qubic's countermeasures reveal how real-world mining campaigns evolve in response to an active ecosystem. Overall, the campaign disrupted Monero without exhibiting a stable profit-maximizing strategy, leaving open disruptive incentives beyond direct mining rewards.
\end{abstract}

\section{Introduction}

Selfish mining can undermine the incentive compatibility of proof-of-work systems~\cite{eyal2014majority}.
By withholding and strategically releasing blocks, a miner can orphan competing blocks and waste other miners' resources.
Despite extensive theoretical study~\cite{Nayak2016StubbornMining,sapirshtein2016optimal}, large-scale deployments have rarely been documented on live networks.

In August 2025, Qubic, a blockchain network built around redirecting mining resources to external computations, launched a high-profile mining campaign on Monero (XMR). Qubic reported selling the mined XMR to fund buybacks and burns of its native QUBIC token and to pay miner bonuses, tying the operation to its own token economy~\cite{qubic2025epoch172,qubic2025takeover}. It publicly described the campaign as a ``51\% takeover'' and a demonstration of selfish mining. The incident therefore raised practical questions that public dashboards could not answer. We ask what occurred on chain, how the campaign was executed operationally, and whether it was economically effective or primarily disruptive.

This framing also raises a question that block-reward profitability alone cannot answer.
Kroll et al. describe a \emph{Goldfinger attack} as an attack whose utility comes from weakening the target cryptocurrency rather than from earning native mining rewards.
We restrict our profitability analysis to mining rewards and revisit the available market-price evidence in Section~\ref{sec:discussion}.

Answering these questions is especially challenging in Monero.
Blocks do not carry explicit pool identifiers, and key signals of selfish mining are off-chain by design.
We therefore study the campaign by combining multiple vantage points: blocks on the main chain and orphan blocks from Monero nodes, Qubic pool job notifications, and Qubic-network artifacts shared by the Monero community.
Table~\ref{tab:research-questions} summarizes the five research questions and our main findings.

\begin{table}[tb]
\centering
\caption{Research questions and main findings.}
\label{tab:research-questions}
\setlength{\tabcolsep}{4pt}
\renewcommand{\arraystretch}{1.08}
\footnotesize
\begin{tabularx}{\linewidth}{L{0.36\linewidth}X}
\toprule
\textbf{Research question} & \textbf{Main finding} \\
\midrule
\textbf{RQ1.} Can Qubic activity be reliably identified despite Monero's privacy design? & Yes. Community-observed data, extra-nonce patterns, pool observations, and disclosed view keys jointly support Qubic labels and interval-level aggregate measurements. \\
\addlinespace[2pt]
\textbf{RQ2.} Did Qubic sustain majority mining power, and what direct chain-level impact did its activity have? & We do not observe sustained majority control. Qubic's share rose sharply in identified periods, but the clearer impact is elevated orphaning and deeper reorganizations. \\
\addlinespace[2pt]
\textbf{RQ3.} Do static selfish mining models explain Qubic's observed main-chain share? & No single fixed strategy explains all periods. Qubic's observed behavior generally fell between standard selfish mining and a more conservative release strategy, while its execution did not appear consistently optimized. \\
\addlinespace[2pt]
\textbf{RQ4.} Did Qubic gain mining rewards from this behavior, and what execution factors shaped gains or losses? & The withholding intervals do not show consistent reward gains over honest mining. Later difficulty-adjustment spillovers mitigated the shortfall, but Qubic remained below the honest mining baseline across the identified periods and subsequent gaps. \\
\addlinespace[2pt]
\textbf{RQ5.} How did the Monero ecosystem respond, and how should those responses affect interpretation of the incident? & The community monitored Qubic's mining activity and relayed reconstructed blocks when possible. Qubic treated this exposure as an operational risk and progressively strengthened its countermeasures, reducing visibility into its internal mining activity over time. \\
\bottomrule
\end{tabularx}
\end{table}

We make four contributions:
\begin{itemize}
  \item \textbf{Multi-source attribution and dataset.} We combine node-observed blocks, Qubic pool signals, and community-observed data, and validate the resulting attribution against disclosed view keys while quantifying its coverage limits.
  \item \textbf{Empirical campaign measurement.} We reconstruct Qubic's mining activity, assess its public claim of majority mining, and measure its effects on orphaning and chain reorganizations.
  \item \textbf{Strategy and economic evaluation.} We compare the observed behavior with the standard selfish mining model and a conservative-release variant, analyze race outcomes, and account for delayed difficulty-adjustment spillovers when evaluating mining rewards.
  \item \textbf{Operational response and inference limits.} We examine community monitoring and Qubic's countermeasures as operational factors that affected block observability and limit what can be inferred about Qubic's intended release strategy.
\end{itemize}

\noindent\textbf{Organization.}
Section~\ref{section:related} reviews related work.
Section~\ref{sec:data-collection} describes the data sources, block attribution, and coverage limits.
Section~\ref{sec:mining-power-analysis} measures Qubic's mining share and its effects on orphaning and chain reorganizations.
Sections~\ref{sec:selfish-mining-analysis}--\ref{sec:timevarying} analyze Qubic's selfish mining behavior and its immediate and delayed effects on mining rewards.
Section~\ref{sec:tactics} analyzes community monitoring, Qubic's countermeasures, and the resulting limits on strategy inference.
Section~\ref{sec:discussion} discusses the broader economic interpretation and mitigation implications, and Section~\ref{sec:conclusion} concludes.

\section{Related Work}
\label{section:related}
\noindent\textbf{Selfish mining and strategy variants.}
Selfish mining was formalized as a profitable deviation under specific network and propagation assumptions~\cite{eyal2014majority}.
Follow-on work generalized the attacker's strategy space, including eclipse variants~\cite{Nayak2016StubbornMining}.
Other work studied optimal and efficiently computable selfish mining strategies using Markov decision processes~\cite{sapirshtein2016optimal,BarZur2020EfficientMDP},
and examined how profitability and thresholds change across protocols such as Ethereum~\cite{FengNiu2019EthereumSelfishMining}. Intermittent selfish mining has also been studied in the context of difficulty adjustment~\cite{Negy2020selfishReExamined}.
Our work builds on these foundations but focuses on what can be validated in a real campaign where internal attacker state is not observable and attribution is difficult.
Recent work has also provided statistical and empirical tools for detecting behavior resembling selfish mining in real PoW systems~\cite{li2024statistical}.

\noindent\textbf{Propagation, forks, and difficulty dynamics.}
Propagation delays and network asymmetries shape stale blocks and tie-breaking, and are central to selfish mining performance~\cite{Decker2013information,Fechner2022PropagationDelays}.
Beyond generic propagation effects, Monero's network-layer attack surface has been studied in the context of eclipse-style isolation~\cite{shi2025eclipse_monero}.
Prior work further showed that difficulty adjustment can interact with variable hash power and adversarial strategies to create unstable throughput or cyclic effects~\cite{Ilie2021UnstableThroughput}.

\noindent\textbf{Defenses and mitigations.}
A diverse set of defense approaches have been proposed, including Fresh Bitcoins~\cite{heilman2014one}, Zeroblock~\cite{solat2017zero}, and Publish-or-Perish style mechanisms~\cite{zhang2017publish}.
Alternative designs such as Fruitchains~\cite{pass2017fruitchains} and StrongChain~\cite{szalachowski2019strongchain} explore fairness and transparency in PoW consensus.
In parallel, operational discussions in the Monero community have proposed pool-level detective mining and Publish-or-Perish-inspired mitigations~\cite{Spagni2025DetectiveMiningIssue140,tevador2025SelfishMiningMitigations}.
Finally, work on Monero mining-pool data publication shows that pool-published data can expose otherwise hidden information, illustrating why operational data can matter for incident validation~\cite{wijaya2021transparency}.

\section{Data Collection and Qubic Block Attribution}
\label{sec:data-collection}

This section describes the data sources and processing steps used in the analysis.
\ifcameraready
\else
See Appendix~\ref{appendix:dataset} for the dataset and analysis code.
\fi

\subsection{Block and mining information collection}

We combine three data sources: Monero block data from nodes, mining job data from Qubic pool job-notification traffic, and Qubic-related block data collected by the Monero community from Qubic network traffic.

\noindent\textbf{Monero block information.} We operated a pruned Monero full node, which validates the blockchain while retaining only a subset of prunable transaction data to reduce storage. This node let us collect main-chain blocks and locally observable orphan blocks that are not available from public explorers. We operated it from September 29 to October 17, 2025. For earlier periods, we queried public Monero full nodes via RPC for historical block and coinbase data. To improve orphan coverage, we probed 20 public nodes, selected the five most responsive nodes, and periodically synchronized their orphan observations. Public-node collection still does not guarantee complete orphan coverage, and some pool-specific orphan blocks may remain unobserved.

\noindent\textbf{Qubic mining pool.}
Qubic distinguishes consensus nodes called \textit{Computors} from miners that receive tasks and return solutions~\cite{qubic_docs_node_types}.
Our observations cover the public pool API and community-collected job and solution traffic, not Qubic's full internal network.
Because we initiated this study after the campaign was already underway, the job-notification data cover its later phase, from September 26 to October 17, 2025.
During this period, we queried the RPC API of the Qubic mining pool at 5-second intervals.
The API has a structure similar to the Stratum protocol, which is widely used by PoW mining pools, and is openly accessible.
Specifically, the \texttt{job\_notify} method returns multiple values, including the mining block height and the previous block hash, enabling miners to obtain the most recent mining jobs.
These records show which block tip Qubic instructed miners to work on.

\noindent\textbf{Monero community observations.}
We also obtained an additional dataset of Qubic-related blocks from Monero community users Sergei Chernykh and DataHoarder, who collected it by monitoring Qubic-related activity. We refer to these records as \emph{community-observed Qubic blocks}. The community used multiple observation channels, including Qubic network traffic and related artifacts, but we do not treat the dataset as a complete Qubic-provided ground truth. We use it as an independent view of Qubic-related blocks, especially orphan blocks that may be missed by Monero-node observation. Merging the community and node-derived datasets by block hash adds 93 community-observed orphan blocks that are absent from the node-derived block table, suggesting limited propagation.

\subsection{Qubic block attribution}
\label{subsec:qubic-attribution}

\noindent\textbf{View-key verification.}
Monero transaction outputs, including coinbase rewards, use one-time addresses that cannot be publicly linked to the recipient's address. Blocks therefore do not reveal which miner controls a reward output. A disclosed view key allows the corresponding reward outputs to be identified retrospectively. Qubic disclosed its view keys only after each weekly epoch ended, so they could not support real-time attribution. This verification was also limited for periods before our node began operating because we could not fully collect historical orphan-block data.
\ifcameraready
\else
Appendix~\ref{appendix:view-key} lists the disclosed view keys used in this study.
\fi

\noindent\textbf{Initial heuristic attribution.}
Before obtaining the community dataset and without using the view keys disclosed by Qubic, we constructed an initial seed set from orphan-fork blocks attributed to Qubic. We examined the extra-nonce area of their coinbase transactions and iteratively refined two structural regex patterns to identify additional candidate blocks in the node data. These patterns extended attribution coverage but were not treated as independent proof of ownership.
\ifcameraready
\else
Appendices~\ref{appendix:extra-nonce-examples} and~\ref{appendix:regex} provide representative examples and the two patterns.
\fi

\noindent\textbf{Cross-source validation and final labels.}
We compared the initial labels with the community dataset over the overlapping observation period. Of the 13,000 initially attributed Qubic blocks, 12,989 also appear in the community dataset, corresponding to a 99.92\% confirmation rate among the initial positive labels. We observe no disagreement in chain or orphan status among the matched blocks. We retain the remaining 11 node-observed orphan candidates because each matches a Qubic extra-nonce pattern and shares the relevant prefix with a verified Qubic main-chain block at the same height. Their coinbase data were not preserved, however, so they cannot be verified directly with a view key.

The community dataset contains 13,771 Qubic blocks in the study period, 782 of which were missed by our initial heuristic attribution. Of these, 689 were present but unlabeled in our pruned-node dataset, while 93 were absent from it. We relabel the former and add the latter community-only orphan blocks. The resulting union contains 58,944 observed blocks, of which 13,782 are attributed to Qubic. Our final labels therefore follow the community-observed hash set, supplemented by the 11 node-observed orphan candidates.

\noindent\textbf{Coverage and parent-block checks.}
The 93 community-only orphans were observable in the Qubic network and verifiable with disclosed view keys, but absent from observations from our Monero node. Excluding them changes the reported quantities only slightly and does not alter our conclusions. Separately, for community-observed Qubic orphans with raw block blobs, we parse the Monero header \texttt{prev\_id} and check whether the parent hash appears in either dataset. This identifies the parent for 1,218 of 1,228 such orphans overall and for 1,213 of 1,216 within our observation window. Because the global node-observed table does not preserve raw blobs or parent hashes for every non-Qubic orphan, this check does not establish complete parent coverage for all observed forks.
\ifcameraready
\else
Appendix~\ref{appendix:fork-linkage-validation} reports the detailed results.
\fi

\section{Qubic's Mining Share and Network Impact}
\label{sec:mining-power-analysis}

This section examines Qubic's mining activity on Monero during the measurement period.
We first quantify its mining power based on attributed blocks, and then examine its impact on orphan blocks and reorganizations.
A focused analysis of Qubic's selfish mining strategy follows in Section~\ref{sec:selfish-mining-analysis}.

\begin{figure}[tbp]
  \centering
  \includegraphics[width=\linewidth]{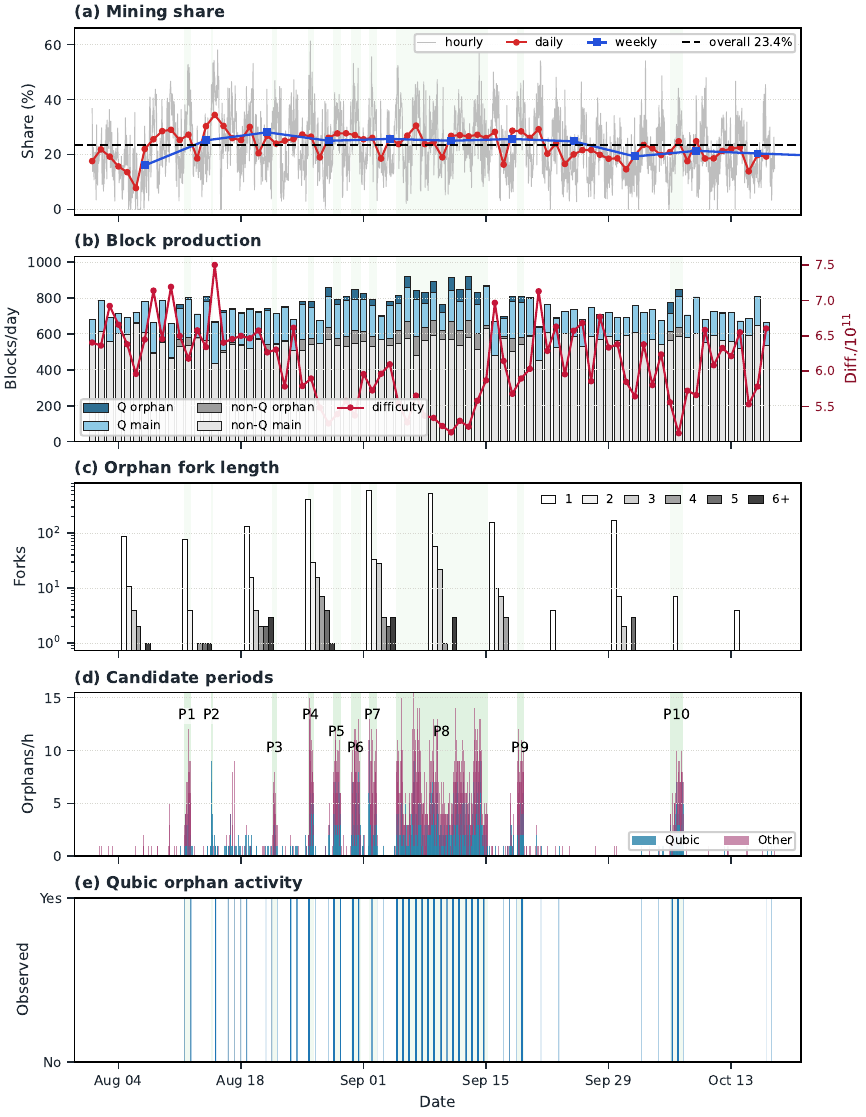}
  \caption{Overview of Qubic's mining activity and selfish mining indicators on a common timeline. Panels show (a) Qubic's observed block-production share, (b) daily block production and average difficulty, (c) weekly orphan-fork length distribution, (d) hourly orphan counts and candidate selfish mining periods, and (e) hours containing observed Qubic orphan blocks. Lightly shaded bands denote the ten candidate periods used in the selfish mining analysis.}
  \label{fig:overview-timeline}
\end{figure}

\subsection{Observed mining share}

Figure~\ref{fig:overview-timeline}(a) shows Qubic's observed block-production share in the Monero network, computed as the ratio of Qubic-attributed blocks to all observed main-chain and orphan blocks over weekly, daily, and hourly windows. Because the pool's physical hashrate is difficult both to measure directly and to reconstruct retrospectively, we use this quantity as an empirical estimate of its relative hashrate $\alpha$ in our selfish mining models.
This calculation uses the combined attribution set described in Section~\ref{subsec:qubic-attribution}, including community-observed Qubic labels.
Across the measurement period, the resulting estimate averages 23.38\%.

A central point in public discussions was Qubic's claim, echoed by several media outlets, that it had achieved a 51\% mining position on Monero~\cite{qubic2025takeover}.
However, the notion of a ``51\% attack'' is often left ambiguous, and our measurements do not support a persistent majority.
In the hourly series, we observe several short intervals where Qubic's share approaches or briefly exceeds 50\%.
In contrast, the corresponding daily and weekly aggregates never reach 51\%, and Qubic does not maintain a stable majority at any point in our dataset.
These results indicate that while Qubic temporarily concentrated substantial mining power, it did not achieve the sustained control typically associated with a practical 51\% attack on the Monero network.

\subsection{Orphaning and reorganizations}

Although Qubic did not sustain majority mining power, periods of elevated Qubic activity coincided with a marked increase in orphaning and deeper forks on Monero.

Figure~\ref{fig:overview-timeline}(b) presents the daily counts of main-chain and orphan blocks, separated into Qubic-attributed and non-Qubic blocks, alongside the average network difficulty.
During periods of elevated Qubic activity, both Qubic and non-Qubic orphan counts increase markedly.
The resulting forks contain blocks from other miners as well as Qubic's own blocks that failed to enter the main chain.

Among the cases for which raw block blobs are available, we directly verify eleven heights at which at least two distinct blocks attributed to Qubic share the same parent.
At nine heights, one Qubic block entered the main chain and another was orphaned.
At two additional heights in P2, both Qubic sibling blocks were orphaned in favor of a non-Qubic block.
Nine of the eleven cases occurred between August 14 and 17, including six in P2.
Mining a second block on the same parent does not extend a private chain, so this behavior wastes the pool's own work and is inconsistent with a rational mining strategy.
We suspect that these events resulted from an early job coordination failure in Qubic's selfish mining operation.
Because our job telemetry begins later, we cannot directly confirm the internal cause.

Figure~\ref{fig:overview-timeline}(c) shows the distribution of orphan fork lengths over time.
Outside periods of elevated Qubic activity, orphan forks are almost exclusively of length one.
During high-activity periods, the distribution shifts toward longer orphan chains and more frequent multi-block forks.
This pattern shows that the campaign coincided with deeper reorganizations rather than only more isolated one-block orphans.

\section{Qubic's Selfish Mining Strategy and Revenue}
\label{sec:selfish-mining-analysis}

We examine how Qubic's selfish mining behavior varied over time and compare its observed mining revenue with model predictions and honest mining baselines.

\subsection{Selfish mining strategies}

Selfish mining, introduced by Eyal and Sirer~\cite{eyal2014majority} and later optimized by Sapirshtein et al.~\cite{sapirshtein2016optimal}, describes how a rational miner (or pool) withholds blocks and selectively publishes a private chain to seek a revenue share exceeding its relative hashrate.
We use the original Eyal--Sirer strategy as our standard selfish mining model.
In practice, miners face uncertainty due to network asynchrony and incomplete information, and may adapt their withholding and release rules accordingly.
Qubic's observed behavior suggests that its withholding and release rules differed from the standard strategy, although its exact internal strategy remains unobservable.

We use the following terminology throughout the analysis.
The attacker's relative hashrate is denoted by $\alpha$.
A \emph{private chain} is a sequence of valid blocks that the attacker has mined but has not yet broadcast.
The \emph{lead} is the number of blocks by which this private chain is ahead of the public chain.
For example, if the public chain is at height 100 and the attacker privately holds blocks up to height 102, the lead is 2.
In the state-machine figures below, state $i$ means that the attacker has lead $i$.
State $0$ means there is no private lead.
State $0'$ is the tie-breaking event that occurs when the attacker publishes a withheld block at the same height as an honest block.
The parameter $\gamma$ is defined only for this state $0'$ tie-breaking event: it is the probability that honest miners who find the next block extend the attacker's branch rather than the competing honest branch.
Thus, $\gamma$ is not an orphan rate and is not estimated from arbitrary non-tip orphan blocks.

Because Qubic's internal block-discovery times and private-chain states are unavailable, we infer its behavior from Monero main-chain and orphan data, Qubic attribution labels, and block timestamps.
We first assess whether the timestamps can approximate event ordering, then use orphan activity to identify sustained selfish mining periods.

\subsection{Reliability of Qubic block timestamps}

We assess whether Qubic's block timestamps approximate discovery and release timing.
To do so independently of the fork patterns analyzed below, we measure the delay between each timestamp and the first subsequent job-fetch record that references the block as its predecessor.

\begin{figure}[tbp]
  \centering
  \begin{minipage}[t]{0.48\linewidth}
    \centering
    \includegraphics[width=\linewidth]{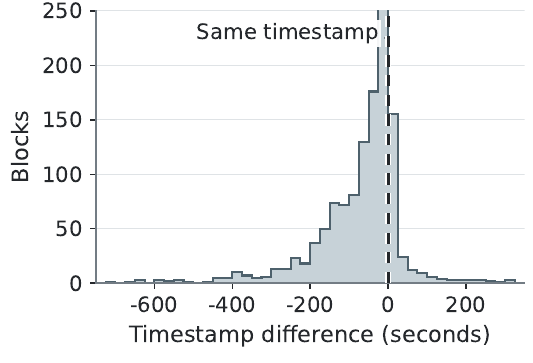}
    \caption{Qubic blocks' timestamp difference distribution with same-height competing blocks, including orphan blocks.}
    \label{fig:timestamp-difference}
  \end{minipage}
  \hfill
  \begin{minipage}[t]{0.48\linewidth}
    \centering
    \includegraphics[width=\linewidth]{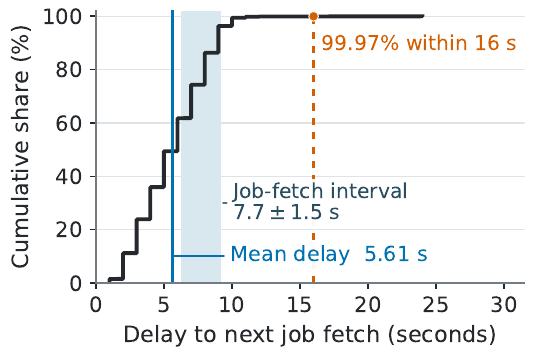}
    \caption{Empirical CDF of the delay between Qubic block timestamps and the first subsequent job-fetch record that references the block. The mean delay is 5.61 seconds. 86.3\% of matched blocks fall within 8 seconds and 99.97\% within 16 seconds.}
    \label{fig:timestamp-distribution}
  \end{minipage}
\end{figure}

The mining client is configured to fetch new jobs approximately every 7 seconds, and the observed average fetch interval in our logs is $7.74 \pm 1.46$ seconds.
The mean delay is 5.61 seconds, with 86.3\% of matched blocks falling within 8 seconds and 99.97\% within 16 seconds (Fig.~\ref{fig:timestamp-distribution}).
This delay distribution is consistent with timestamps being generated within Qubic's normal mining workflow.
Although the comparison cannot rule out selective manipulation, we find no evidence of systematic timestamp shifts large enough to invalidate relative temporal comparisons.
Additionally, community observers who monitored Qubic-network traffic reported no indications of systematic timestamp manipulation.\footnote{Personal communication with Monero community observers.}

Using these timestamps, Fig.~\ref{fig:timestamp-difference} shows that most Qubic blocks competing at the same height have timestamps earlier than or comparable to those of their competitors.
This timing pattern supports early block discovery followed by selective withholding.
The small number of unusually late timestamps occurs predominantly among orphaned Qubic blocks.

We therefore use Qubic's timestamps, together with orphan dynamics, as approximate indicators of relative event ordering and selfish mining activity.
This interpretation assumes that Qubic did not systematically alter timestamps to conceal private-chain lead lengths or release decisions.

\subsection{Periods of sustained selfish mining activity}

Qubic's strategy may vary over time, and selfish mining, if present, is unlikely to be applied uniformly across the entire observation window.
To separate periods of sustained abnormal orphan activity from the rest of the observation window, we use the heuristic in Alg.~\ref{alg:heuristic}.
We treat such activity as an indicator of potential selfish mining.

\begin{algorithm}
\caption{Selfish Mining Period Heuristic}
\label{alg:heuristic}
\begin{algorithmic}[1]
\Require Blocks $B$ with timestamps and orphan flags, thresholds $\tau_{\min}$, $d_{\min}$, $g_{\max}$
\Ensure Valid period spans $M$

\State \textbf{Step 1:} Construct the complete hourly series
\State $H \gets \text{complete hourly range covered by } B$
\ForAll{$h \in H$}
\State $C[h] \gets |\{b \in B : b.\text{is\_orphan} \land \lfloor b.\text{timestamp} \rfloor_{\text{hour}} = h\}|$
\EndFor

\State \textbf{Step 2:} Find maximal contiguous segments
\State $S \gets \text{maximal segments of } \{h \in H : C[h] \geq \tau_{\min}\}$
\State $S \gets \{(s_i,e_i) \in S : |e_i-s_i| \geq d_{\min}\}$

\State \textbf{Step 3:} Merge segments with gaps $\leq g_{\max}$ hours
\State $M \gets \text{merge}(S, g_{\max})$

\State \Return $M$
\end{algorithmic}
\end{algorithm}

The heuristic operates in three steps.
First, we construct a complete hourly series over the observation window and compute the orphan count $C[h]$ for each hour, assigning zero to hours with no observed orphan blocks.
Second, we identify maximal contiguous segments in which every hour satisfies a minimum orphan-count threshold $\tau_{\min}$ and the total segment length exceeds a minimum duration $d_{\min}$.
Third, we merge neighboring segments separated by gaps shorter than $g_{\max}$ hours to tolerate brief fluctuations in activity.
The result is a set $M$ of candidate selfish mining periods.

\begin{figure}[tb]
    \centering
    \includegraphics[width=\linewidth]{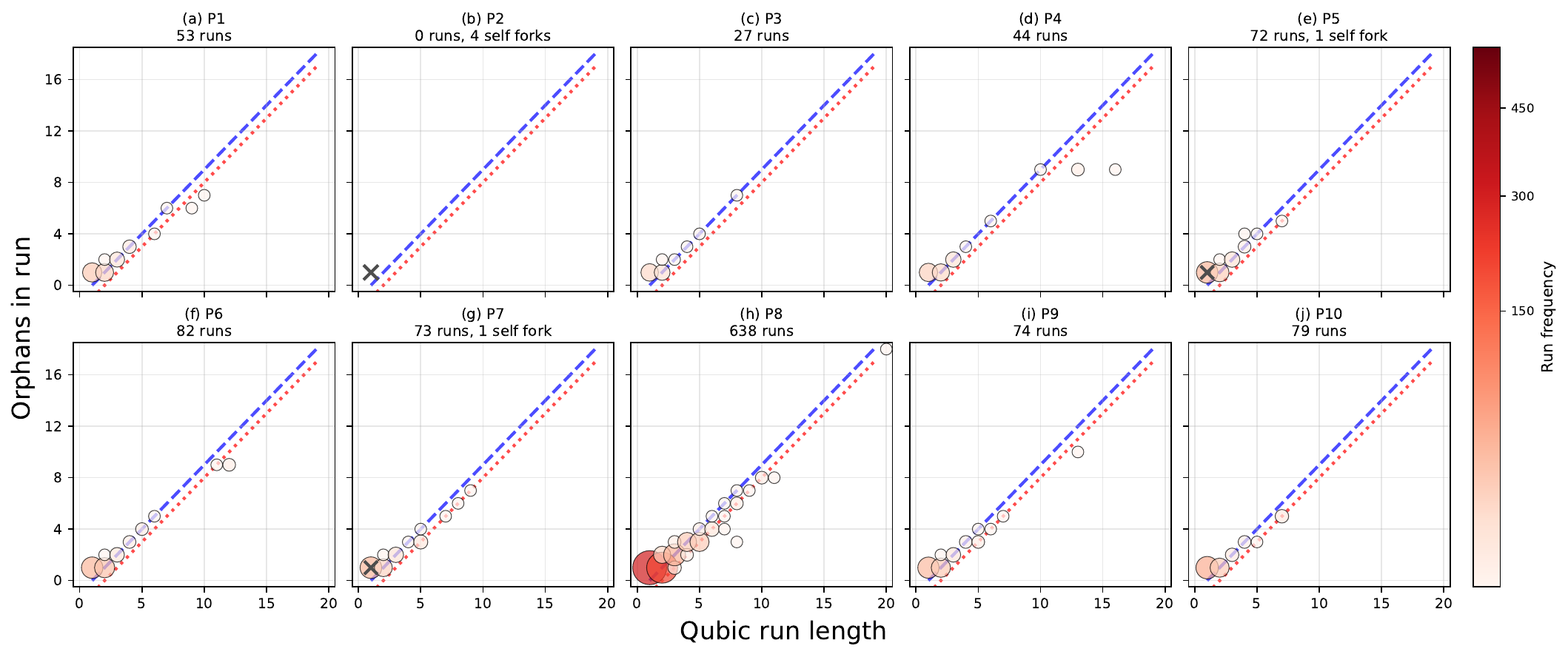}
    \caption{Distribution of orphan blocks per Qubic run length across ten candidate periods, shown in panels (a) through (j). Red circles visualize the frequency of fork runs involving Qubic and other miners. Gray crosses mark Qubic self forks associated with an accepted Qubic block, which are excluded from inference about Qubic's release strategy because they do not identify a private chain lead. Same-parent Qubic pairs in which both blocks were orphaned fall outside this run-based plot. Reference lines show $y=x-1$ (blue dashed) and $y=x-2$ (red dotted).}
    \label{fig:run_by_periods}
\end{figure}

For Fig.~\ref{fig:run_by_periods}, we distinguish fork runs involving Qubic and other miners from Qubic self forks in which two blocks attributed to Qubic share the same parent and height.
The latter do not identify a private chain lead and are excluded from inference about Qubic's release strategy.

In our evaluation, we set $\tau_{\min} = 2$ orphan blocks per hour, $d_{\min} = 4$ hours, and $g_{\max} = 6$ hours.
These settings capture sustained orphan activity while preventing short gaps from splitting a continuous span of activity into multiple periods.
Our heuristic captures sustained, observable phases of Qubic's selfish mining operation and may miss brief or sporadic activity.
The resulting P1--P10 candidate periods form the basis of the focused analyses that follow.
To check robustness to the threshold choices, we repeat the analysis over all 80 parameter combinations.
In every setting, Qubic's target-rate-normalized yield remains below the corresponding estimated $\alpha$, so the aggregate underperformance is not specific to our default parameters.
\ifcameraready
\else
Appendix~\ref{appendix:threshold-sensitivity} provides the detailed sensitivity analysis, and the artifact includes the full grid.
\fi

Figure~\ref{fig:overview-timeline}(d) shows the resulting candidate periods on the timeline, while Fig.~\ref{fig:overview-timeline}(e) shows the hours in which Qubic orphan blocks were observed.
Applying this heuristic yields ten periods (P1--P10) during which Qubic's block share and involvement in orphan blocks are noticeably higher than their global averages.
Qubic's estimated share averages 28.33\% across these periods, compared with 23.38\% over the full measurement window.
This association suggests that sustained, observable selfish mining activity was more common when Qubic's estimated share was relatively high.

Figure~\ref{fig:run_by_periods} further illustrates how Qubic's behavior varies across the identified periods by plotting the lengths of runs controlled by Qubic against the number of associated orphan blocks. In P1, P3, and P4, most runs that provide release information lie close to the blue line ($y=x-1$), corresponding to release at lead~1. P2 instead contains four plotted Qubic self forks, two additional same-parent Qubic pairs that were both orphaned, and no run that provides release information.

During P8, where Qubic's activity is most intensive, we observe many runs aligned with patterns indicative of release at lead~2, suggesting a more conservative release strategy that avoids exposing the pool to tie situations when the lead is only one block. This shift supports the view that Qubic adapted its release rule over time, motivating the need for an analytical model that captures this conservative release behavior.

\subsection{Standard and conservative selfish mining models}
\label{subsec:analytical-model}

To interpret the observed periods, we compare them against analytical models of selfish mining.
We first recall the standard selfish mining revenue function $R_{\text{selfish}}(\alpha,\gamma)$, which gives the pool's expected main-chain revenue share as a function of its relative hashrate $\alpha$ and the tie-breaking parameter $\gamma$.
Figure~\ref{fig:selfish_mining_state_machine} shows the corresponding state machine.
When the selfish miner has a private lead, it keeps mining privately. When the honest network catches up, the selfish miner publishes withheld blocks to create or win a race.
In state $0'$, the race can resolve in three ways.
An honest miner may extend the honest branch with probability $(1-\alpha)(1-\gamma)$, an honest miner may extend the selfish branch with probability $(1-\alpha)\gamma$, or the selfish miner may find the next block with probability $\alpha$.
Equation~\ref{eq:selfish-mining} gives the Eyal--Sirer revenue expression~\cite{eyal2014majority}.

\begin{equation} \label{eq:selfish-mining}
R_{\text{selfish}}(\alpha,\gamma) %
= \frac{\alpha (1 - \alpha)^2 (4\alpha + \gamma (1 - 2\alpha)) - \alpha^3}{1 - \alpha (1 + (2 - \alpha)\alpha)}
\end{equation}

We then consider a conservative-release variant of selfish mining, which we call the conservative strategy.
The conservative strategy follows the Eyal--Sirer state transitions except at private lead 3, as shown in Fig.~\ref{fig:modified_selfish_mining_state_machine}.
If the honest network finds a block in that state, the attacker fully publishes its three-block private chain and returns to state 0 instead of retaining a private lead.
This additional transition represents a more conservative release decision motivated by Qubic's observed behavior.
We compare Qubic's observations with both models without assuming that either model captures its exact time-varying strategy.

\begin{figure}[t]
  \centering
  \begin{minipage}[t]{0.45\textwidth}
    \vspace{0pt}
    \centering
    \resizebox{\linewidth}{!}{%
    \begin{tikzpicture}[
    >=Stealth,
    semithick,
    start chain=main going right,
    node distance=0.7cm,
    mynode/.style={
      draw,
      ellipse,
      minimum width=11mm,
      minimum height=7mm,
      inner sep=1pt
    },
    every loop/.append style={-latex}
]

\coordinate[on chain] (s0-pos);
\node[mynode, at=(s0-pos), yshift=-9mm] (s0) {$0$};
\node[mynode, on chain] (s1) {$1$};
\node[mynode, on chain] (s2) {$2$};
\node[mynode, on chain] (s3) {$3$};
\node[mynode, on chain] (s4) {$4$};
\node[on chain] (ellipsis) {$\dots$};

\node[mynode,above=12mm of s0] (sp) {$0'$};

\path[-latex] (sp.south west) edge[bend right=40]
    node[left, align=left] {$(1-\gamma)(1-\alpha)$ \\ $\gamma(1-\alpha)$ \\ $\alpha$} (s0.120);
\path[-latex] (sp.south)      edge (s0.north);
\path[-latex] (sp.south east) edge[bend left=40] (s0.60);

\path[-latex] (s1.north) edge[bend right=45] node[above,pos=.6,yshift=2pt] {$1-\alpha$} (sp.east);

\path[-latex] (s0) edge[loop left,looseness=7] node[below,yshift=-2pt] {$1-\alpha$} (s0);

\path[-latex] (s0.north east) edge[bend right=25] node[below] {$\alpha$} (s1.south west);

\path[-latex] (s2.south west) edge[bend left=30] node[below,pos=.5,yshift=-2pt] {$1-\alpha$} (s0.south);

\path[-latex] (s1.north east) edge[bend left=25] node[above] {$\alpha$} (s2.north west);
\path[-latex] (s2.north east) edge[bend left=25] node[above] {$\alpha$} (s3.north west);
\path[-latex] (s3.north east) edge[bend left=25] node[above] {$\alpha$} (s4.north west);

\path[-latex] (s3.south west) edge[bend left=25,draw=classicalStrategy,line width=0.9pt]
    node[below,text=classicalStrategy] {$1-\alpha$} (s2.south east);
\path[-latex] (s4.south west) edge[bend left=25] node[below] {$1-\alpha$} (s3.south east);

\path[-latex,opacity=0] (s3.south west) edge[bend left=30]
    node[below,pos=.5,yshift=-2pt] {$1-\alpha$} (s0.south);

\path[-latex] (s4.north east) edge[bend left=25]  node[above] {$\alpha$} (ellipsis.north west);
\path[-latex] (ellipsis.south west) edge[bend left=25] node[below] {$1-\alpha$} (s4.south east);

\end{tikzpicture}
    }
    \captionof{figure}{Standard selfish mining state machine. State $i$ is the attacker's private lead, and state $0'$ denotes the tie-breaking event.}
    \label{fig:selfish_mining_state_machine}
  \end{minipage}
  \hfill
  \begin{minipage}[t]{0.45\textwidth}
    \vspace{0pt}
    \centering
    \resizebox{\linewidth}{!}{%
    \begin{tikzpicture}[
    >=Stealth,
    semithick,
    start chain=main going right,
    node distance=0.7cm,
    mynode/.style={
      draw,
      ellipse,
      minimum width=11mm,
      minimum height=7mm,
      inner sep=1pt
    },
    every loop/.append style={-latex}
]

\coordinate[on chain] (s0-pos);
\node[mynode, at=(s0-pos), yshift=-9mm] (s0) {$0$};
\node[mynode, on chain] (s1) {$1$};
\node[mynode, on chain] (s2) {$2$};
\node[mynode, on chain] (s3) {$3$};
\node[mynode, on chain] (s4) {$4$};
\node[on chain] (ellipsis) {$\dots$};

\node[mynode,above=12mm of s0] (sp) {$0'$};

\path[-latex] (sp.south west) edge[bend right=40]
    node[left, align=left] {$(1-\gamma)(1-\alpha)$ \\ $\gamma(1-\alpha)$ \\ $\alpha$} (s0.120);
\path[-latex] (sp.south)      edge (s0.north);
\path[-latex] (sp.south east) edge[bend left=40] (s0.60);

\path[-latex] (s1.north) edge[bend right=45] node[above,pos=.6,yshift=2pt] {$1-\alpha$} (sp.east);

\path[-latex] (s0) edge[loop left,looseness=7] node[below,yshift=-2pt] {$1-\alpha$} (s0);

\path[-latex] (s0.north east) edge[bend right=25] node[below] {$\alpha$} (s1.south west);

\path[-latex] (s2.south west) edge[bend left=30] node[below,pos=.5,yshift=-2pt] {$1-\alpha$} (s0.700);

\path[-latex] (s3.south west) edge[bend left=30,draw=conservativeStrategy,line width=0.9pt]
    node[below,pos=.5,yshift=-2pt,text=conservativeStrategy] {$1-\alpha$} (s0.south);

\path[-latex] (s1.north east) edge[bend left=25] node[above] {$\alpha$} (s2.north west);
\path[-latex] (s2.north east) edge[bend left=25] node[above] {$\alpha$} (s3.north west);
\path[-latex] (s3.north east) edge[bend left=25] node[above] {$\alpha$} (s4.north west);

\path[-latex] (s4.south west) edge[bend left=25] node[below] {$1-\alpha$} (s3.south east);

\path[-latex] (s4.north east) edge[bend left=25]  node[above] {$\alpha$} (ellipsis.north west);
\path[-latex] (ellipsis.south west) edge[bend left=25] node[below] {$1-\alpha$} (s4.south east);

\end{tikzpicture}
    }
    \captionof{figure}{State machine for the conservative strategy. At lead 3, an honest discovery triggers full publication and a return to state 0.}
    \label{fig:modified_selfish_mining_state_machine}
  \end{minipage}
\end{figure}

Using this model, we derive a closed-form expression for the expected revenue under the conservative strategy as a function of $\alpha$ and $\gamma$.

\begin{proposition}
\label{prop:modified_selfish_revenue}
Let $R_{\mathrm{mod}}(\alpha,\gamma)$ denote the selfish pool's long-run
fraction of accepted blocks under the conservative strategy encoded by
the state machine in Fig.~\ref{fig:modified_selfish_mining_state_machine}, where $\alpha \in (0,\tfrac12)$ is the pool's
relative hashrate and $\gamma \in [0,1]$ is the tie-breaking parameter.
Then
\begin{equation}
\label{eq:modified-selfish-mining}
R_{\mathrm{mod}}(\alpha,\gamma)
=
\frac{
\alpha\big(
-2\alpha^{3}\gamma
+ 3\alpha^{3}
+ 5\alpha^{2}\gamma
- 9\alpha^{2}
- 4\alpha\gamma
+ 4\alpha
+ \gamma
\big)
}{
1 - \alpha - 2\alpha^{2} + \alpha^{3} - \alpha^{4}
}.
\end{equation}
\end{proposition}

\begin{proof}
\ifcameraready
\else
See Appendix~\ref{appendix:proof}.
\fi
\end{proof}

\begin{figure}[tbp]
  \centering
  \includegraphics[width=0.80\linewidth]{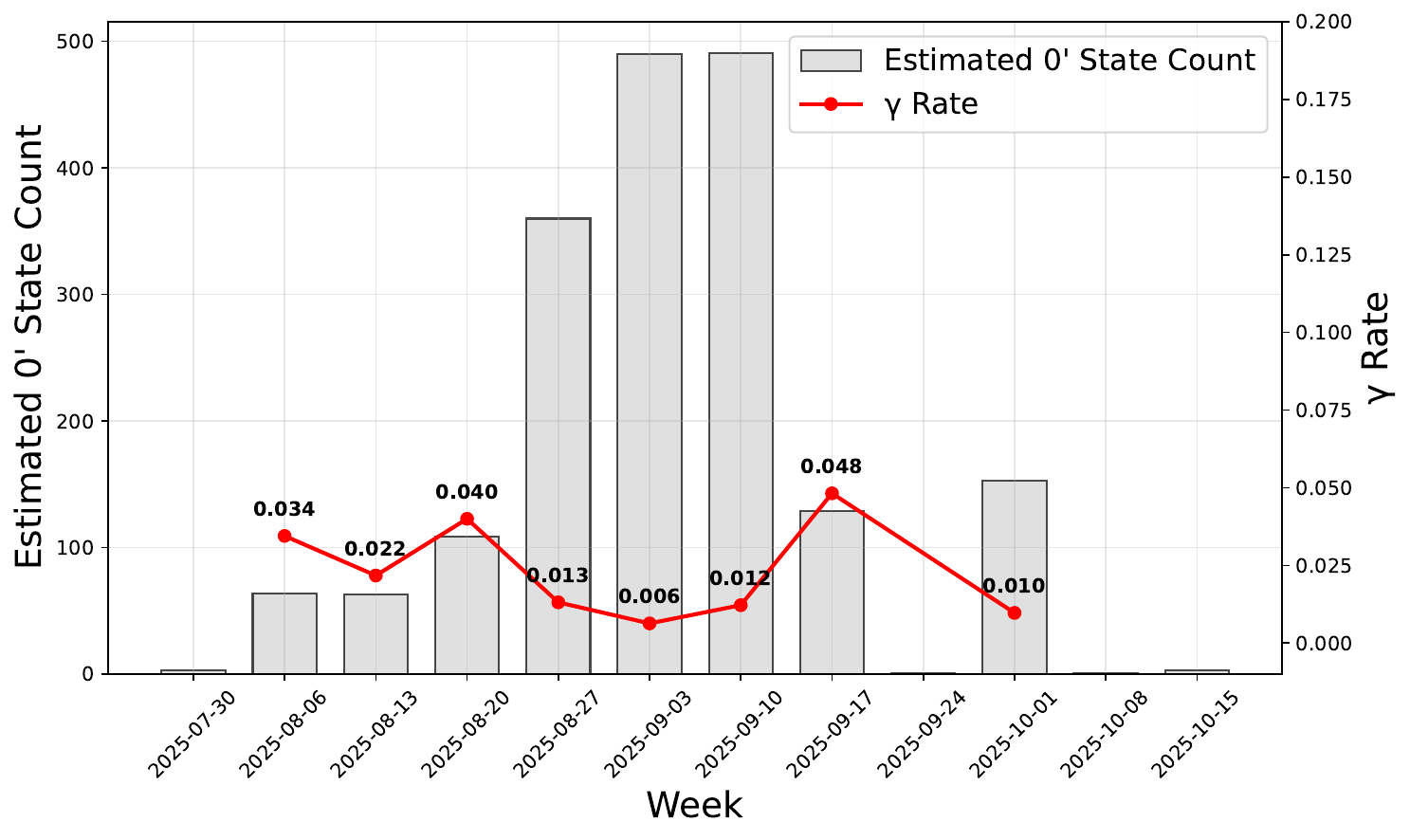}
  \caption{Weekly estimates of the tie-breaking parameter $\gamma$ and the number of observed state $0'$ tie-breaking events. We use only resolved races containing exactly one Qubic block and a competing block attributed to another miner. Heights containing multiple Qubic blocks and non-tip orphan blocks are excluded because they do not reveal a binary tie-breaking choice.}
  \label{fig:gamma}
\end{figure}

\subsection{Model predictions and observed revenue}

We now apply the analytical model using parameters inferred from the identified selfish mining periods.
For each period, we estimate Qubic's relative hashrate $\alpha$ from its share of main-chain and orphaned blocks.
We estimate $\gamma$ only from resolved state $0'$ tie-breaking events: cases where exactly one Qubic block competes with a block attributed to another miner at the same height and the next main-chain block reveals which branch the network extended.
If the next block is mined by an honest miner and extends Qubic's branch, we count it as a $\gamma$ success. If it extends the honest branch, we count it as a failure.
We exclude non-tip orphan blocks from this calculation, because they may be discovered after the race has already been resolved and therefore do not measure tie-breaking behavior.
Figure~\ref{fig:gamma} shows that the weekly estimates of $\gamma$ are close to zero.
We attribute this to delayed propagation because Qubic blocks tend to arrive at nodes after competing blocks despite having earlier timestamps.

\begin{figure}[tb]
    \centering
    \includegraphics[width=0.92\linewidth]{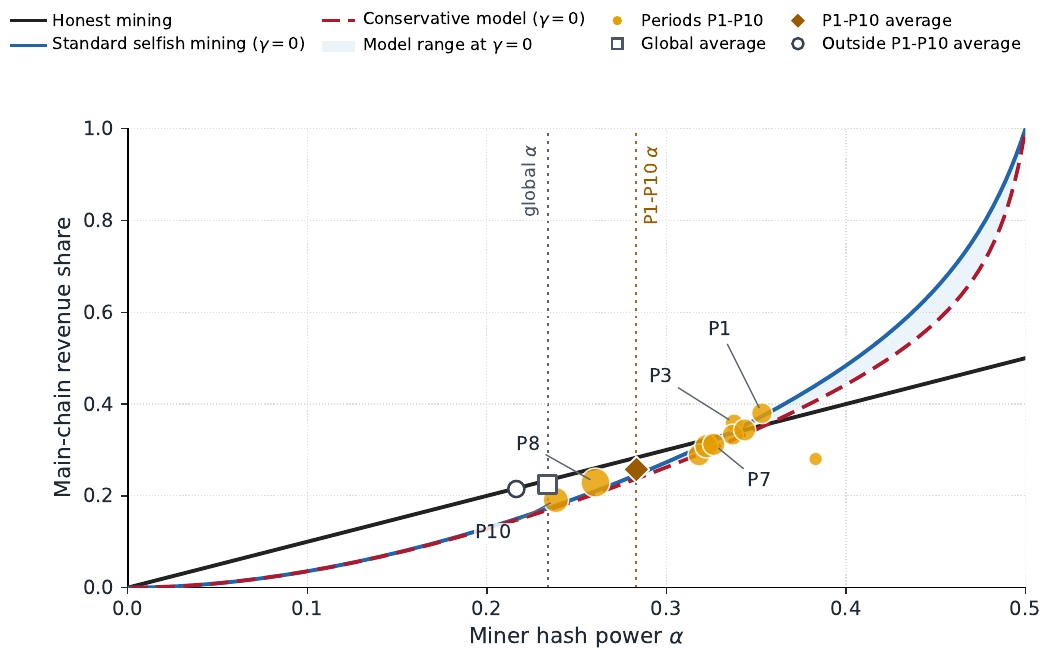}
    \caption{Theoretical main-chain revenue shares predicted by the standard and conservative models at $\gamma \approx 0$, compared with Qubic's observed main-chain shares. The shaded area shows the revenue range obtained by mixing the two fixed strategies in different proportions. Period markers show P1--P10, with marker size proportional to period scale. The filled diamond shows the P1--P10 average, while hollow markers show the global average and the average outside P1--P10.}
    \label{fig:selfish_mining_theory}
\end{figure}

At the average hashrate share of $28.33\%$ during the identified selfish mining periods and $\gamma \approx 0$, the standard selfish mining model yields an expected revenue ratio of $R_{\text{selfish}} \approx 24.56\%$, while the conservative model yields $R_{\text{mod}} \approx 23.71\%$.
Both are below the honest mining baseline at the same $\alpha$.
Qubic's P1--P10 main-chain share is $25.75\%$, also below the honest mining baseline.
The same relation holds without P8, with a main-chain share of $29.88\%$ against an estimated $\alpha$ of $31.53\%$.
Separately, its share of the blocks expected at Monero's target rate is $25.11\%$, a $3.22$ percentage-point shortfall from $\alpha$ that measures the active-period economic opportunity cost.
The net contribution of transaction fees was below 3\% of realized block rewards and did not materially affect our conclusions.

The period-level points, however, do not align cleanly with either reference-model prediction.
Some periods lie closer to the standard model, some are closer to the conservative model, and P1 and P3 outperform the honest baseline despite the aggregate underperformance.
\ifcameraready
\else
Table~\ref{tab:observed_revenue} reports the detailed period values.
\fi
The next section examines observed tie-breaking outcomes as one direct reason why period-level shares can depart from the corresponding $\gamma\approx0$ model predictions.

\Needspace{6\baselineskip}
\section{Race Outcomes and Limits of Static Models}
\label{sec:analysis-difference}

The preceding comparison treats the standard and conservative models as fixed reference strategies.
We next examine Qubic's outcomes in state $0'$ tie-breaking events as one directly observable source of deviation.

The reference-model predictions use the global estimate $\gamma\approx0$.
Local race outcomes can nevertheless differ across periods, and Qubic sometimes won observed state $0'$ tie-breaking events at rates above its period-level mining share.
These are cases where exactly one Qubic block competes with a block attributed to another miner at the same height and the next accepted block reveals which branch the network extended.
We exclude heights containing multiple Qubic blocks because these self forks do not represent a binary tie-breaking event.

\begin{figure}[tbp]
    \centering
    \includegraphics[width=0.72\linewidth]{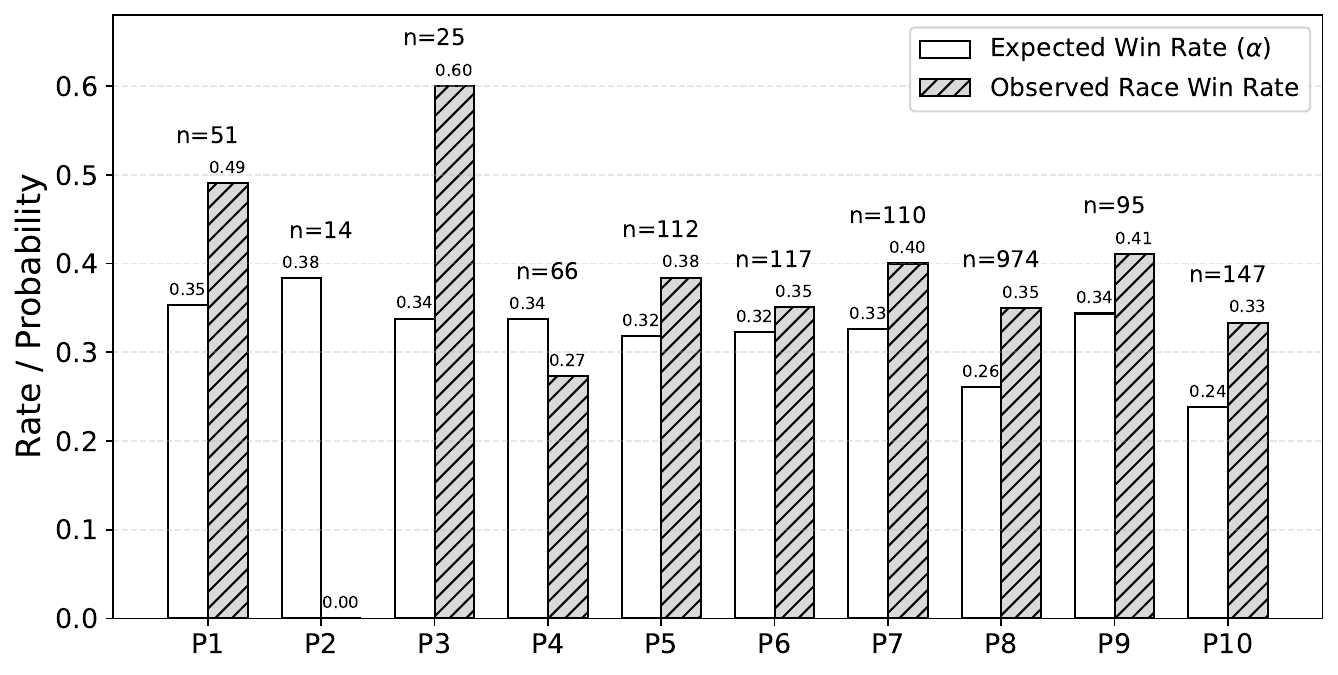}
    \caption{Qubic's winning ratio in observed state $0'$ tie-breaking events. The reference win probability is $\alpha+(1-\alpha)\gamma$, which reduces to the period-level block share $\alpha$ under the global estimate $\gamma\approx0$. The observed rate counts events in which the next main-chain block extends Qubic's branch. Heights containing multiple Qubic blocks are excluded.}
    \label{fig:race_resolution}
\end{figure}

As illustrated in Fig.~\ref{fig:race_resolution}, Qubic's winning rate in these race conditions was higher than its period-level estimated share $\alpha$ in eight of the ten periods.
For example, P1 has an observed race winning rate of $0.49$ against $\alpha=0.35$, P3 has $0.60$ against $\alpha=0.34$, and P7 has $0.40$ against $\alpha=0.33$.

\Needspace{5\baselineskip}
By contrast, P2 recorded no Qubic-branch wins ($0/14$) and contained six parent-linked Qubic self forks within roughly 95 minutes, suggesting temporary operational instability.
P4 also remained below $\alpha$.
More broadly, these results show that period-average $\alpha$ does not fully capture the local conditions under which Qubic released withheld blocks.

These local advantages can help explain why some observed main-chain shares exceed the corresponding $\gamma\approx0$ model predictions.
They do not explain the aggregate target-rate shortfall or establish consistently profitable execution.
Race counts are limited in some periods, and high win rates can coexist with orphan losses and missed honest mining rewards.
The next section therefore examines whether lower difficulty following the identified periods increased Qubic's rewards during the subsequent gaps and offset its active-period losses.

\section{Intermittent Mining and Difficulty-Adjustment Spillovers}
\label{sec:timevarying}

Qubic earned fewer blocks than the honest mining baseline in most of the identified selfish mining periods.
Evaluating only these periods may miss a delayed benefit because slower main-chain growth can lower Monero's sliding-window difficulty in the intervals that follow.
Intermittent selfish mining has been studied as a difficulty-adjustment phenomenon~\cite{Negy2020selfishReExamined}.
To capture both the immediate and delayed effects, we pair each identified period $P_k$ with its following gap $G_k$ and analyze them as a single accounting sequence.

\subsection{Accounting model}
\label{subsec:daa-accounting}

Each $G_k$ begins after $P_k$ and ends when the next identified period starts.
For each $P_k$, we define the signed balance relative to honest mining as
\begin{equation}
\Delta_{\mathrm{active}}(P_k)
=
B_Q^{\mathrm{main}}(P_k)
-
\alpha_k N_{\mathrm{target}}(P_k),
\label{eq:active-balance}
\end{equation}
where $N_{\mathrm{target}}(P_k)$ is the number of blocks expected at Monero's target rate.
A negative value indicates that Qubic obtained fewer accepted blocks than expected from honest mining at its estimated hashrate.
This target-rate accounting expresses the period's effect in block-equivalents, allowing direct comparison with the subsequent DAA effect.

\subsection{Estimating delayed effects on gap rewards}
\label{subsec:gap-replay}

If selfish mining reduces main-chain production during $P_k$, Monero's sliding-window DAA can leave the following gap with lower difficulty.
To estimate this delayed effect, we add synthetic main-chain slots to restore target-rate production during each $P_k$, then replay Monero's sliding-window DAA over the reconstructed timeline.
This replay estimates the difficulty path that would have followed if main-chain production during $P_k$ had remained at the target rate.
We then compare the observed difficulty at each Qubic-attributed accepted block in $G_k$ with the corresponding replayed difficulty.
\ifcameraready
\else
Appendix~\ref{appendix:difficulty-discussion} discusses the time-scale implications of Monero's sliding-window DAA.
\fi

Let $D_i$ be the observed difficulty at a Qubic-attributed accepted block in $G_k$, and let $\widehat{D}_i$ be the corresponding replayed difficulty.
The block-equivalent DAA spillover is
\begin{equation}
\Delta_{\mathrm{DAA}}(G_k)
=
\sum_{i\in Q(G_k)}
\left(1-\frac{D_i}{\widehat{D}_i}\right),
\label{eq:daa-windfall}
\end{equation}
where $Q(G_k)$ is the set of Qubic-attributed accepted blocks in the gap.
If the observed difficulty is lower than the replayed difficulty, the block contributes a positive fractional block-equivalent.

\begin{figure}[tb]
    \centering
    \includegraphics[width=0.90\linewidth]{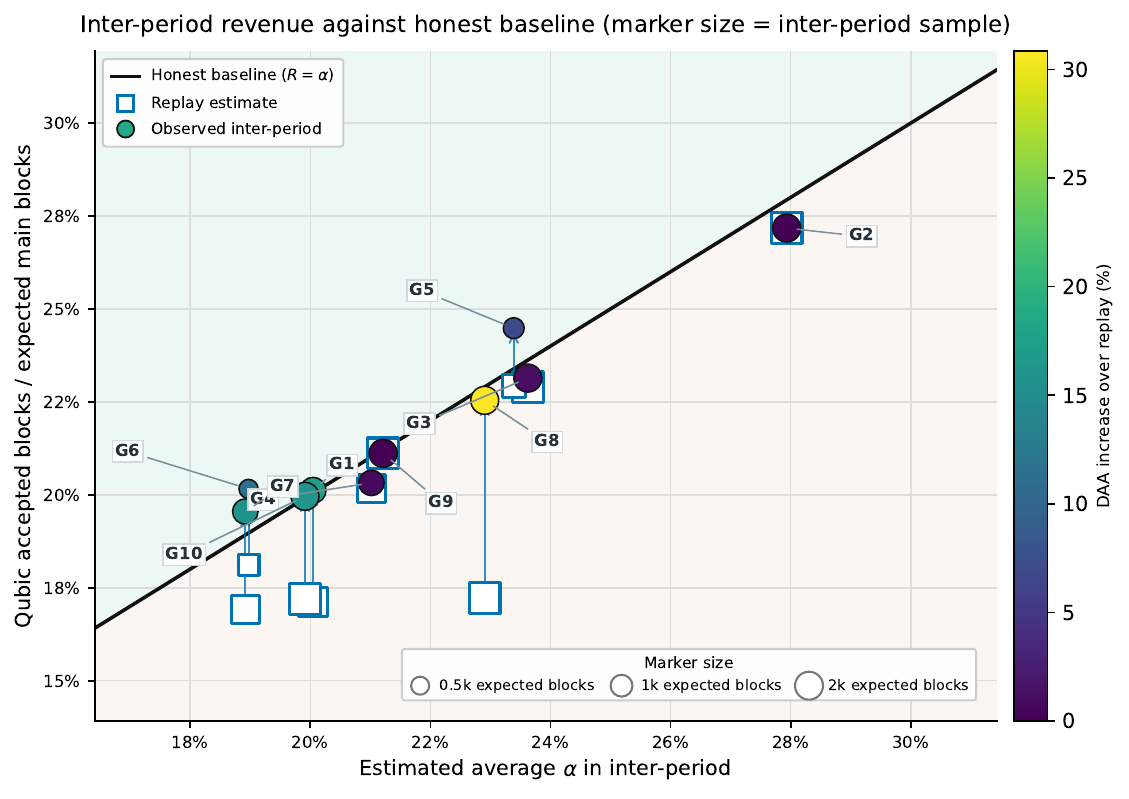}
    \caption{Qubic's gap revenue under the observed and replayed difficulty paths. Each $G_k$ denotes the gap after identified period $P_k$. The black line is the honest mining baseline $R = \alpha$. Hollow squares show the replay estimate, filled circles show observed gap revenue, and vertical arrows show the DAA spillover. Marker size encodes gap sample size, capped at the 2k expected-block guide for readability.}
    \label{fig:interperiod-daa-spillover}
\end{figure}

\subsection{Empirical spillover}
\label{subsec:daa-results}

Figure~\ref{fig:interperiod-daa-spillover} compares Qubic's observed gap rewards with the replay estimates.
We measure the DAA effect as the increase from the replay estimate to the observed outcome.
This comparison is distinct from whether the observed point lies above the honest mining line.
Across all gaps, Qubic obtained $7{,}806$ accepted blocks, while the replay yields $7{,}344.2$ block-equivalents.
The difference is $+461.8$ block-equivalents, or $+6.3\%$ relative to the replay estimate.
Against a target-rate honest mining expectation of $7{,}855.3$ blocks, the observed gap total remains $49.3$ blocks lower.
Summed across P1--P10, the period balance is $-410.7$ block-equivalents.
Across the identified periods and subsequent gaps, Qubic obtained $11{,}045$ accepted blocks against a target-rate honest mining expectation of $11{,}505.0$, a shortfall of $460.0$ block-equivalents, or $4.0\%$.

P8 and its following gap provide a representative example.
During P8, Qubic's estimated hashrate was $26.05\%$.
It obtained $1{,}700$ accepted blocks, compared with an honest mining expectation of $1{,}961.5$, giving $\Delta_{\mathrm{active}}(P_8)=-261.5$ block-equivalents.
During $G_8$ (P8$\rightarrow$P9), Qubic obtained $541$ accepted blocks, while the replay estimate was $413.5$ block-equivalents.
The resulting $\Delta_{\mathrm{DAA}}(G_8)=+127.5$ block-equivalents substantially raised gap rewards.
However, the target-rate honest mining expectation for $G_8$ was $549.7$ blocks, leaving the gap $8.7$ blocks lower and the P8--$G_8$ pair $270.2$ block-equivalents below the corresponding honest mining baselines.

The estimated DAA spillover increased Qubic's rewards in the later gaps.
Even with this increase, its combined reward across the identified periods and subsequent gaps remained $4.0\%$ below the target-rate honest mining baseline.

\section{Community Monitoring and Qubic Countermeasures}
\label{sec:tactics}

The preceding sections analyzed Qubic's campaign on Monero quantitatively, using observed data and theoretical mining models. The campaign, however, was a real-world operation in which one blockchain project targeted another live network, with active communities on both sides. The Monero community responded quickly by monitoring Qubic's mining activity and relaying observations and reconstructed blocks to Monero nodes. Qubic, in turn, introduced countermeasures. This section examines how that interaction affected block observability and informs our interpretation of Qubic's observed release behavior.

\subsection{Community monitoring and block relay}

To detect Qubic's private mining activity, Monero community members deployed monitoring nodes across the network.
By connecting to Qubic's low-level mining network from multiple observation nodes, collecting packet-level data, and analyzing local logs, these community observers reconstructed the timing of mining jobs and their corresponding solution messages on Qubic's \textit{Computor} network.

Each job encodes a block template, a candidate block assembled for mining that specifies the header and the set of transactions.
This allowed the community observers to track which template Qubic used at a given time and when the corresponding solution was observed.
They checked the origin and integrity of the collected messages using the digital signatures of Qubic's dispatcher and \textit{computor} nodes and matched each job to its corresponding solution.
They then used the PoW hash in the Result field of each solution message to match the solution to a reconstructed Monero block.
Reception-log timestamps provided the timing information.

When these records were sufficient to reconstruct a valid Monero block, community members attempted to propagate it before Qubic's intended release.
The relay was not integrated with a protocol-level mechanism for coordinating Monero nodes, making it difficult to translate these efforts into a reliable chain-level response.

The processed records also helped the community identify Qubic-related blocks and assemble the dataset used in our analysis.
We assessed the resulting dataset against observable block patterns and Qubic's later-disclosed view keys, obtaining the strong agreement reported in Section~\ref{subsec:qubic-attribution}.

\subsection{Qubic countermeasures}

Qubic nevertheless appears to have regarded the possibility of early private-block exposure as an operational risk and soon introduced countermeasures.\footnote{The operational details in this subsection are based primarily on personal communication with Monero community observers.}
First, to prevent leakage of mining-solution data, it introduced encryption for solution messages circulating on its \textit{Computor} network.
Initially, Qubic applied a simple XOR-based scheme.
Because the messages contained fixed or readily inferable fields, we infer that the key could be recovered through known-plaintext analysis.
Community observers also reported recovering the key.
Qubic then changed its encryption method and keys multiple times, rotated keys regularly, and adjusted message-routing paths, making real-time observation increasingly difficult.
These measures reduced the community's access to solution data, although the open \textit{Computor}-network architecture continued to expose some activity to external observers.
Over time, Qubic's network appeared to shift toward a more controlled and centralized configuration, with tighter control over task dissemination and increasing reliance on a small set of dispatching points.

As a more fundamental countermeasure, Qubic began withholding transactions from block templates shared on its network.
The missing transactions were revealed only when Qubic later published or relayed the block, preventing observers from reconstructing the complete Monero block in real time.
Once the missing transactions became available, community observers reconstructed past blocks and relayed them for retrospective verification.
Because this reconstruction occurred only after disclosure, it did not provide a timely mechanism for influencing chain selection.

\subsection{Operational impact and inference limits}

This sequence shows that Qubic's operation was shaped by more than hashrate dynamics.
The community's relay efforts were not part of a coordinated Monero protocol response, so their direct effect on chain selection or Qubic's realized rewards cannot be established.
Qubic responded with encryption, routing changes, and withheld transactions.
These countermeasures progressively reduced the observability of its later activity.

Observed orphan lengths and release patterns may therefore reflect a mixture of Qubic's intended strategy, network propagation, monitoring pressure, and Qubic's own countermeasures.
Our analysis describes the behavior visible in the data rather than reconstructing Qubic's exact internal strategy.
The reference strategies therefore serve as benchmarks for interpreting release behavior that changed over time, not as exact descriptions of individual forks.

\section{Discussion}
\label{sec:discussion}

\noindent\textbf{Economic interpretation.}
Qubic's campaign may not have been motivated solely by mining revenue.
In the Goldfinger model, an attacker may benefit by undermining confidence in the target cryptocurrency~\cite{kroll2013economics}.
Qubic's promotional framing and token-linked reward structure suggest incentives beyond Monero block rewards~\cite{qubic2025epoch172,qubic2025takeover}.
However, daily CoinGecko prices do not indicate a lasting effect on Monero.
From August 1 to October 17, XMR ranged from \$235.58 to \$343.44 and ended at \$293.15, 1.0\% below its initial price.
QUBIC ranged from USD~$1.38\times10^{-6}$ to USD~$3.27\times10^{-6}$ and ended at USD~$1.47\times10^{-6}$, 43.2\% below its initial price.
Price data alone cannot determine whether the campaign produced broader benefits.
The attention it brought to Qubic may itself have provided promotional value beyond mining rewards.

\noindent\textbf{Mitigation implications.}
Freshness-aware chain-selection rules, including Publish-or-Perish-style mechanisms discussed after the campaign, could reduce the value of late withheld blocks~\cite{heilman2014one,solat2017zero,zhang2017publish,tevador2025SelfishMiningMitigations}.
Their effectiveness depends on assumptions about timing and propagation.
Detective mining could use information exposed by a public selfish mining pool to mine on its leading private branch~\cite{Spagni2025DetectiveMiningIssue140,lee2023rethinking}.
In practice, transaction withholding creates operational risks, and intentionally extending the attacker's branch raises social concerns.
Both responses address specific tactics, while systems with modest and concentrated hash power remain exposed because of limits in decentralization, propagation, and observability.

\section{Conclusion}
\label{sec:conclusion}

Qubic's campaign neither sustained majority control nor consistently followed a textbook selfish mining strategy.
It increased orphaning and reorganization depth.
Nevertheless, across the identified periods and subsequent gaps, Qubic remained 4.0\% below the target-rate honest mining baseline despite occasional tie-breaking advantages and later DAA spillovers.
Community monitoring and Qubic's countermeasures further show that real-world campaigns evolve with changing observability and operational responses.
Evaluating such campaigns requires considering their on-chain effects, underlying incentives, public claims, and ecosystem responses.

\bibliographystyle{plainurl}
\bibliography{references}

\ifcameraready
\else
  \clearpage
  \appendix
  \section{Dataset and analysis code}
\label{appendix:dataset}
The datasets, analysis code, and numerical results underlying the figures and tables are available at \url{https://github.com/shlee-lab/Qubic-selfish-mining-study}.

\section{Representative Extra-Nonce Examples}
\label{appendix:extra-nonce-examples}

\begin{table}[H]
\centering
\caption{Examples of \texttt{extra\_nonce} values in coinbase transactions from Qubic and non-Qubic pools.}
\label{tab:extra-nonce-pattern}
\setlength{\tabcolsep}{4pt}
\renewcommand{\arraystretch}{1.1}
\begin{adjustbox}{width=\linewidth,center}
\begin{tabular}{@{}>{\ttfamily}l@{\hspace{6pt}}>{\ttfamily}l@{}}
\multicolumn{1}{c}{\textbf{Blocks by Qubic pool}} &
\multicolumn{1}{c}{\textbf{Blocks by Other pools}} \\
\hline
\multicolumn{1}{c}{\texttt{extra\_nonce}} &
\multicolumn{1}{c}{\texttt{extra\_nonce}} \\
\hline
\rowcolor[HTML]{FFF0E5} a18300008f031173362951280000000000 &
\cellcolor[HTML]{E6F6E6}3d5bf9d77da9ba00000000000000000000 \\
\rowcolor[HTML]{FFF0E5} \texttt{8e8300008f0311735d0637400000000000} & \cellcolor[HTML]{E6F6E6}\texttt{0000000000000001337bcdecc400000000000000000000000000000000000000} \\
\rowcolor[HTML]{FFF0E5} \texttt{718300008f031173a00100000000000000} & \cellcolor[HTML]{E6F6E6}\texttt{00000003cd754c00000000000000000000} \\
\rowcolor[HTML]{FFF0E5} \texttt{638300008f0311733f0200000000000000} & \cellcolor[HTML]{E6F6E6}\texttt{f8b08baa} \\
\rowcolor[HTML]{FFF0E5} \texttt{5a8300008f0311734f0c0b280000000000} & \cellcolor[HTML]{E6F6E6}\texttt{00000000000000001829d427bb00000000000000000000000000000000000000} \\
\rowcolor[HTML]{FFF0E5} \texttt{ed8200008f031173c2271b300000000000} & \cellcolor[HTML]{E6F6E6}\texttt{000000000000000065c66da4e10000000100000cb70000000000000000000000} \\
\end{tabular}
\end{adjustbox}
\end{table}

\section{Qubic's regex patterns}
\label{appendix:regex}

\lstset{basicstyle=\ttfamily,breaklines=true}

\begin{lstlisting}[
  caption={Qubic extra-nonce regex patterns},
  label={lst:qubic-extra-nonce-regex},
  frame=single,
  breaklines=true
]
# Pattern A
([0-9a-f]{4})0000([0-9a-f]{8})([0-9a-f]{8})0{10}
# Pattern B
([0-9a-f]{4})0100([0-9a-f]{8})([0-9a-f]{8})0{10}
\end{lstlisting}

\section{List of Qubic's view keys}
\label{appendix:view-key}

The view keys listed in Table~\ref{table:qubic-viewkeys} were obtained from the official Qubic Discord server, accessible via \url{https://qubic.org/}. They were disclosed by Qubic's lead developer, \textit{dkat}. These keys let the community retrospectively check ownership of Qubic-mined blocks.

\begingroup
\footnotesize
\renewcommand{\arraystretch}{1.05}
\setlength{\tabcolsep}{3pt}
\begin{longtable}{@{}L{0.16\textwidth} L{0.46\textwidth} L{0.32\textwidth}@{}}
\caption{Qubic view key list}\label{table:qubic-viewkeys}\\
\toprule
\textbf{Date / Epoch} & \textbf{Reward address} & \textbf{View key} \\
\midrule
\endfirsthead
\caption[]{Qubic view key list (continued)}\\
\toprule
\textbf{Date / Epoch} & \textbf{Reward address} & \textbf{View key} \\
\midrule
\endhead
\midrule
\multicolumn{3}{r@{}}{\scriptsize Continued on the next page}\\
\endfoot
\bottomrule
\endlastfoot
May 27, 2025 &
{\ttfamily\seqsplit{47hhGMKbWpKfxDiqcejWGicVvQHEYd45AEaUyKVjcZywL8c8mtjN3oACGfdrsLrPGP2r49gvTBnBiTVQcEkfBNFEKCDy7ME}} &
{\ttfamily\seqsplit{577fd4a7278f55d2a9230d32823b81497b2e854d4a8702b1256a17cda42a760d}} \\
\addlinespace
Aug.\ 6, 2025\newline epoch 172 &
{\ttfamily\seqsplit{43oMtdwB5aaCuM9vVaiY6u7XgxCGLwA563C7b5V3oSTSjDdhiBkWeGxeZZSuD4wAydMzbvNWrF9iRGmwoMnhYnMTMcZjBrv}} &
{\ttfamily\seqsplit{e935552c5665117a6ecc9fbbfd4156595c75774014606130a01003720e063201}} \\
\addlinespace
Aug.\ 13, 2025\newline epoch 173 &
{\ttfamily\seqsplit{49upGQgCYzxMKfBU9hYe8QH3fQQMSdReiAJx3vo9bq6K4YegbgP39rVKNNGh9tA3VobYMkyGxvDc1J9FnVFw8f4UT7BsDhf}} &
{\ttfamily\seqsplit{0b21ef509769c6d95899cca7ccd86b89333ca4ce0dfa3fd5aa304059aed0f903}} \\
\addlinespace
Aug.\ 20, 2025\newline epoch 174 &
{\ttfamily\seqsplit{4AqzG7scWP19yNsFuJNpQ2CNF7LGxJNtJaEennAA48KbLX6a7PTazW4c3FTwBPfjJ4TFq3xpZhvGvgygyVCtuXxSLWrAk3S}} &
{\ttfamily\seqsplit{05ada241eea8b262241762cb6be291be3aaf1237560a0ddd1fd4ea5cf502120f}} \\
\addlinespace
Aug.\ 28, 2025\newline epoch 175 &
{\ttfamily\seqsplit{45hzuq7TBR3J89EXkAmZuqYDGbpdCxf5qXU2dMYb5jorR2JuS7V4T9XhuMAGwF7CG895Suf6XR4PUWbLUhB5UnzcS3d6MDy}} &
{\ttfamily\seqsplit{9267d1762b0f3262029be73e30f5158159c2f38e86b9d745231e57141afccd0a}} \\
\addlinespace
Sep.\ 3, 2025\newline epoch 176 &
{\ttfamily\seqsplit{49heVqhSznN9eoatkooNJLHHsRV6XiitDeW4J92cgney8BFfuacZGmzSA3fRKEHooC7X9xzCP9VXN6uK7XrpoXF35FXfQCP}} &
{\ttfamily\seqsplit{d761a707408f9693f9a453501dc04df6c92b82309a90f23884564dabfed70106}} \\
\addlinespace
Sep.\ 10, 2025\newline epoch 177 &
{\ttfamily\seqsplit{42Vt47oLyRT7C1Ch3BbapKFZgs5Hip5m3RrRVT2dbjDT8NWs76gc77NgfZvzXpZnPYGgVZFf79T5TSKWSjFxYWk4A77WGa6}} &
{\ttfamily\seqsplit{91c313b9cb0cc45e03e2f6f97e9d61566f8d0636b4bb3bf59c59022972caad09}} \\
\addlinespace
Sep.\ 17, 2025\newline epoch 178 &
{\ttfamily\seqsplit{47GwPhLcnWshcbekVshrzxJZwXXfDRUrjb7T6CfR1HaaiokeBxwAsQnFp779bF6rW43giviWwYbsoT1KehsGnP5L7v1vuF6}} &
{\ttfamily\seqsplit{c9f5d5027465f4ff51538210e4fa110756e956c064897f969c4a60863e227f0c}} \\
\addlinespace
Sep.\ 24, 2025\newline epoch 179 &
{\ttfamily\seqsplit{44UsmtpAE5GC8U8vnLp7FqUfAYkWL5YYZJLNFdQrb46ePGpSH58ydJ2QtfmEgR834AQphJYwsLVnJRrE1uFhT38bQnTebXm}} &
{\ttfamily\seqsplit{a5e32d32ea8d1aac9ed47b7679ecf2cc4884dc5b388d9b539fede7ae5389f603}} \\
\addlinespace
Oct.\ 1, 2025\newline epoch 180 &
{\ttfamily\seqsplit{42sk9bcpJeVQRkZvZu6bidP2WJ6w6QR15cEEcBBAJHhJ4sA1aviq2NDK6JLHSneTxCLWpQkoHBiDN7hPiMiH5WWaS1VyM9T}} &
{\ttfamily\seqsplit{5b78ba1a936efe94acb8e13fce72ee3581267ef668a9c0f8967883ab12394602}} \\
\addlinespace
Oct.\ 8, 2025\newline epoch 181 &
{\ttfamily\seqsplit{48PSv1UrxcrQY6m1ZaDMXSMt38kEvZNhjWNgiSmJoUL7BFjm5A4XkiBKt2ApF5ydqsDtaMfZK8WBT7PtabvKGMfZUdZqjUe}} &
{\ttfamily\seqsplit{1fbd6085b25183aadd0c241e94adb4379bfb145cd51b968c7bf068d171275902}} \\
\addlinespace
Oct.\ 15, 2025\newline epoch 182$^{*}$ &
{\ttfamily\seqsplit{48BZD46hnvGJmh5kn4py9oEupM8uL3ZobK59GDMWkZBdHyB2ALiSD6rGa2u9inZgMtegfKciaanDYNDFEw8oGHKmAcYEQTo}} &
{\ttfamily\seqsplit{938e19ee3f4fc0c2515d5b1dc509d2f450f7e6d27a06d3197db761a4e8405809}} \\
\addlinespace
Oct.\ 15, 2025\newline epoch 182$^{*}$ &
{\ttfamily\seqsplit{42y5h2KPKhKTW82xqf7XHMFUz32Hs8ubgCowd8Q4y5RY8XqvNCEFzX2cCexyfsLtdD1BjT5mRMDHiBrC1t8CaT1RDeDsMrg}} &
{\ttfamily\seqsplit{173ad08b17faf1a672ee8314c79e4d2e931ed85210c925d722cb400a2c7c3405}} \\
\addlinespace
Oct.\ 23, 2025\newline epoch 183 &
{\ttfamily\seqsplit{45w3hfgjzJjHiDSsVKx4nKdHawTGrWZr4WPT5LL4qeNx4fyyRQ73cNiLGGrrJ5pDjP8LHeDJfTMCs1UN7eBpWTPy1YX7YCV}} &
{\ttfamily\seqsplit{726d15c6d9963e41965d87e311f1e51ff5722badfa824336e893fb01a11acb00}} \\
\addlinespace
Oct.\ 29, 2025\newline epoch 184 &
{\ttfamily\seqsplit{48CzXZX9YkcTTKAP8qc2cMfFiXpLQpxDDVEDNn5mTrF5aX9thAY7eqfUzJkwqkXtaZhb7Ggv9rjCJYeYRrZfHKPJF2BbDtx}} &
{\ttfamily\seqsplit{7b8f7098b538d92756b0f5e81f2677caec13f9d02ec71ab4feec06ca642da90e}} \\
\addlinespace
Nov.\ 6, 2025\newline epoch 185 &
{\ttfamily\seqsplit{4A6mfADoDhNEAodQChmXRTSgsK4gheMna73TebsmKw4rRggore7U1p8No32pJytUPTfoS5xk11aDh93BzgqZphbkUykELdm}} &
{\ttfamily\seqsplit{d696c09c6a95b91fbf709711f027797c9bcee4a3d73d86e0361e95126655fe0e}} \\
\end{longtable}
\endgroup

\noindent{\scriptsize $^{*}$ Epoch 182 is the only epoch in which Qubic used two reward addresses.}

\section{Parent-block validation}
\label{appendix:fork-linkage-validation}

For some community-observed Qubic orphans, the dataset contains raw Monero block blobs.
We parse the block-header \texttt{prev\_id} from these blobs and check whether the parent hash is present in either the node-observed block table or the community-observed dataset.
Table~\ref{tab:fork-linkage-validation} summarizes the resulting parent-linkage coverage.
This validation is scoped to community-observed Qubic orphans with raw block blobs. The global node-observed block table does not retain raw blobs or parent hashes for every orphan block.

\begin{table}[H]
\centering
\caption{Parent-linkage validation for community-observed Qubic orphans.}
\label{tab:fork-linkage-validation}
\setlength{\tabcolsep}{5pt}
\renewcommand{\arraystretch}{1.08}
\begin{adjustbox}{max width=\textwidth}
\begin{tabular}{lrrrrrr}
\toprule
\textbf{Scope} & \textbf{Community Qubic orphans} & \textbf{Parent known} & \textbf{Coverage} &
\textbf{Parent in nodes} & \textbf{Parent in community set} & \textbf{Missing} \\
\midrule
All such orphans & 1,228 & 1,218 & 99.19\% & 1,209 & 251 & 10 \\
Orphans in observation window & 1,216 & 1,213 & 99.75\% & 1,209 & 246 & 3 \\
Orphans also observed by nodes & 1,123 & 1,123 & 100.00\% & 1,123 & 220 & 0 \\
\bottomrule
\end{tabular}
\end{adjustbox}
\end{table}

\section{Qubic's Observed Reward Metrics by Period}

Table~\ref{tab:observed_revenue} reports the period-level quantities used in Section~\ref{sec:selfish-mining-analysis}.
Estimated $\alpha$ is Qubic's share among observed main-chain and orphan blocks.
Target yield divides the number of accepted Qubic blocks by the expected number of main-chain blocks at the target rate.
Main-chain share instead uses the actual number of accepted main-chain blocks and is therefore comparable with the stationary revenue models.
The count columns distinguish all accepted main-chain blocks, all Qubic-attributed blocks, and accepted Qubic blocks.

\begin{center}
    \centering
    \captionof{table}{Qubic's observed reward metrics}
    \label{tab:observed_revenue}
        \begin{adjustbox}{width=\linewidth}
        \begin{tabular}{lrrrrrr}
            \toprule
            \textbf{Period} & \textbf{Estimated $\alpha$} & \textbf{Target Yield} & \textbf{Main-chain Share} & \textbf{Main Chain} & \textbf{Qubic Total} & \textbf{Qubic Main} \\
            \midrule
            P1  & 0.3532 & 0.3386 & 0.3799 & 508  & 219  & 193 \\
            P2  & 0.3831 & 0.3083 & 0.2803 & 132  & 59   & 37  \\
            P3  & 0.3378 & 0.3545 & 0.3589 & 326  & 127  & 117 \\
            P4  & 0.3370 & 0.3314 & 0.3333 & 507  & 217  & 169 \\
            P5  & 0.3182 & 0.2778 & 0.2893 & 605  & 245  & 175 \\
            P6  & 0.3223 & 0.2926 & 0.3082 & 769  & 313  & 237 \\
            P7  & 0.3264 & 0.2870 & 0.3123 & 634  & 265  & 198 \\
            P8  & 0.2605 & 0.2258 & 0.2288 & 7429 & 2333 & 1700 \\
            P9  & 0.3436 & 0.3533 & 0.3436 & 617  & 268  & 212 \\
            P10 & 0.2384 & 0.1811 & 0.1909 & 1053 & 299  & 201 \\
            \midrule
            \textbf{Avg (P1-P10)}   & \textbf{0.2833} & \textbf{0.2511} & \textbf{0.2575} & \textbf{12580} & \textbf{4345} & \textbf{3239} \\
            \textbf{Global Average} & \textbf{0.2338} & \textbf{0.2236} & \textbf{0.2244} & \textbf{55937} & \textbf{13782} & \textbf{12555} \\
            \bottomrule
        \end{tabular}
    \end{adjustbox}
\end{center}

Across P1--P10, the aggregate estimated $\alpha$ is 28.33\%, while the target-rate-normalized yield is 25.11\%, a difference of $-3.22$ percentage points from the honest mining baseline.
The corresponding main-chain share is 25.75\%, or $-2.58$ percentage points relative to $\alpha$.
Using main-chain share, P1 and P3 exceed their respective estimated $\alpha$, P9 is approximately equal to it, and the remaining periods are below it.

\section{Threshold sensitivity for period detection}
\label{appendix:threshold-sensitivity}

We evaluate the period-detection heuristic over $\tau_{\min}\in\{1,2,3,4\}$, $d_{\min}\in\{2,4,6,8\}$ hours, and $g_{\max}\in\{2,4,6,8,12\}$ hours.
Target-rate-normalized yield follows the definition used in the active-period economic analysis: Qubic main-chain blocks divided by the number of expected main-chain blocks over the detected period duration.
For each of the 80 parameter combinations, we rerun period detection and recompute the number and duration of the detected periods, $\alpha$, target-rate-normalized yield, and overlap with the default periods.
The yield remains below the corresponding $\alpha$ in every configuration.
Table~\ref{tab:threshold-sensitivity-summary} summarizes groups of configurations, Table~\ref{tab:threshold-sensitivity-selected} reports representative settings, and the artifact provides the full 80-point grid.

\begin{center}
\centering
\captionof{table}{Aggregate sensitivity of period detection to threshold choices. Values are medians over each grid subset.}
\label{tab:threshold-sensitivity-summary}
\setlength{\tabcolsep}{6pt}
\renewcommand{\arraystretch}{1.05}
\begin{tabular}{lrrrrr}
\toprule
\textbf{Grid} & \textbf{N} & \textbf{Periods} & \textbf{$\alpha$} & \textbf{Target yield} & \textbf{Yield $-\alpha$} \\
\midrule
Baseline & 1 & 10 & 28.33\% & 25.11\% & -3.22pp \\
Near baseline & 12 & 10 & 28.72\% & 25.59\% & -3.13pp \\
$\tau_{\min}\ge2$ & 60 & 12 & 28.92\% & 25.82\% & -3.10pp \\
All grid & 80 & 12 & 28.60\% & 25.51\% & -3.10pp \\
\bottomrule
\end{tabular}
\end{center}

\begin{center}
\centering
\captionof{table}{Representative threshold-sensitivity settings for selfish mining period detection.}
\label{tab:threshold-sensitivity-selected}
\setlength{\tabcolsep}{5pt}
\renewcommand{\arraystretch}{1.08}
\begin{adjustbox}{width=\textwidth}
\begin{tabular}{lrrrrrrrrr}
\toprule
\textbf{Setting} & \textbf{$\tau_{\min}$} & \textbf{$d_{\min}$} & \textbf{$g_{\max}$} & \textbf{Periods} &
\textbf{Duration} & \textbf{$\alpha$} & \textbf{Target yield} & \textbf{Yield $-\alpha$} & \textbf{Baseline recall} \\
\midrule
Loose threshold & 1 & 4h & 6h & 14 & 476h & 28.35\% & 25.42\% & -2.93pp & 100.00\% \\
Short duration & 2 & 2h & 6h & 13 & 450h & 28.19\% & 24.90\% & -3.29pp & 100.00\% \\
Small merge gap & 2 & 4h & 2h & 11 & 425h & 28.53\% & 25.30\% & -3.23pp & 98.84\% \\
Baseline & 2 & 4h & 6h & 10 & 430h & 28.33\% & 25.11\% & -3.22pp & 100.00\% \\
Long duration & 2 & 8h & 6h & 10 & 411h & 28.65\% & 25.57\% & -3.08pp & 95.58\% \\
Higher threshold & 3 & 4h & 6h & 10 & 391h & 28.82\% & 25.68\% & -3.15pp & 90.93\% \\
Strict threshold & 4 & 4h & 6h & 16 & 302h & 31.51\% & 28.85\% & -2.66pp & 70.23\% \\
\bottomrule
\end{tabular}
\end{adjustbox}
\end{center}

\section{Time-scale and difficulty adjustment discussion}
\label{appendix:difficulty-discussion}

This time-scale issue differs across protocols. In Bitcoin, difficulty adjustments occur over long epochs of 2016 blocks, or about two weeks, so the averaging window is large. An attacker who reduces main-chain efficiency via selfish mining may need to absorb the revenue impact for a long duration before difficulty responds. This can make intermittent attacks less attractive unless the attacker is willing to tolerate extended shortfall.

Monero updates difficulty frequently using a sliding-window estimator, so it responds more quickly when main-chain block production slows. The attacker may therefore face a shorter period of reduced rewards before difficulty decreases, making it more attractive to alternate between selfish and honest mining.

\section{Proof of Proposition~\ref{prop:modified_selfish_revenue}}
\label{appendix:proof}

\begin{proof}
Let $\beta=1-\alpha$, and let $\pi_s$ be the stationary probability of
state $s$ in Fig.~\ref{fig:modified_selfish_mining_state_machine}.

\medskip
\noindent\textbf{Stationary probabilities.}
The balance equations for states $0'$, $1$, and $2$ follow directly
from the state machine:
\begin{equation}
\pi_{0'}=\beta\pi_1,\qquad
\pi_1=\alpha\pi_0,\qquad
\pi_2=\alpha\pi_1.
\label{eq:modified-low-state-balance}
\end{equation}
In particular, the figure has a transition from state $2$ to state $0$,
not from state $2$ to state $1$.

For each $i\ge2$, consider the cut between
$S_i=\{0,0',1,\ldots,i\}$ and its complement. In stationarity, the
probability flows in both directions across this cut are equal. The
only outward transition is $i\to i+1$, with flow $\alpha\pi_i$.
For $i\ge3$, the only inward transition is $i+1\to i$, with flow
$\beta\pi_{i+1}$. For $i=2$, it is the full-publication transition
$3\to0$, also with flow $\beta\pi_3$. Therefore,
\begin{equation}
\alpha\pi_i=\beta\pi_{i+1}
\qquad (i\ge2).
\label{eq:modified-cut-balance}
\end{equation}
Combining \eqref{eq:modified-cut-balance} with
\eqref{eq:modified-low-state-balance} gives
\begin{equation}
\pi_1=\alpha\pi_0,\qquad
\pi_{0'}=\alpha\beta\pi_0,\qquad
\pi_i=\frac{\alpha^i}{\beta^{i-2}}\pi_0\quad(i\ge2).
\label{eq:modified-stationary-probabilities}
\end{equation}

These relative weights also satisfy the remaining balance equation at
state $0$, since
\[
\beta\pi_0+\pi_{0'}+\beta\pi_2+\beta\pi_3
=
(\beta+\alpha\beta+\alpha^2\beta+\alpha^3)\pi_0
=\pi_0.
\]
Moreover, $\alpha<1/2$ implies $\alpha/\beta<1$, so the geometric tail
is summable:
\[
\sum_{i=3}^{\infty}\pi_i
=
\frac{\alpha^3}{1-2\alpha}\pi_0.
\]
After normalization, these weights define the unique stationary
distribution of the irreducible chain. The normalization factor is not
needed because it cancels from the revenue ratio.

\medskip
\noindent\textbf{Expected accepted blocks.}
Let $E_s$ be the expected number of accepted selfish-miner blocks per
block-finding event, and let $E_t$ be the corresponding number of all
accepted blocks.

At state $0$, an honest block contributes one accepted block to $E_t$.
At state $0'$, the next block resolves the race and finalizes two blocks.
The selfish miner's expected share of those blocks is
$2\alpha+\gamma\beta$.
At state $2$, an honest block causes the selfish miner to publish two
blocks and return to state $0$.
At state $3$, an honest block causes full publication of three blocks
and a return to state $0$.
Finally, at every state $i\ge4$, an honest block causes one selfish
block to be accepted while the private lead decreases to $i-1$.
Thus
\begin{align}
E_s
&=
\pi_{0'}(2\alpha+\gamma\beta)
+2\beta\pi_2
+3\beta\pi_3
+\beta\sum_{i=4}^{\infty}\pi_i,
\label{eq:modified-attacker-reward}\\
E_t
&=
\beta\pi_0
+2\pi_{0'}
+2\beta\pi_2
+3\beta\pi_3
+\beta\sum_{i=4}^{\infty}\pi_i.
\label{eq:modified-total-reward}
\end{align}

From \eqref{eq:modified-stationary-probabilities},
\[
\sum_{i=4}^{\infty}\pi_i
=
\frac{\alpha^4}{\beta(1-2\alpha)}\pi_0.
\]
Substituting the stationary probabilities into
\eqref{eq:modified-attacker-reward} and
\eqref{eq:modified-total-reward} gives
\begin{align*}
\frac{E_s}{\pi_0}
&=
\alpha\beta(2\alpha+\gamma\beta)
+2\beta\alpha^2
+3\alpha^3
+\frac{\alpha^4}{1-2\alpha},\\
\frac{E_t}{\pi_0}
&=
\beta
+2\alpha\beta
+2\beta\alpha^2
+3\alpha^3
+\frac{\alpha^4}{1-2\alpha}.
\end{align*}
Multiplying both expressions by $1-2\alpha$ and collecting terms yields
\begin{align*}
(1-2\alpha)\frac{E_s}{\pi_0}
&=
\alpha\bigl(
-2\alpha^3\gamma
+3\alpha^3
+5\alpha^2\gamma
-9\alpha^2
-4\alpha\gamma
+4\alpha
+\gamma
\bigr),\\
(1-2\alpha)\frac{E_t}{\pi_0}
&=
1-\alpha-2\alpha^2+\alpha^3-\alpha^4.
\end{align*}
The long-run revenue ratio is $R_{\mathrm{mod}}=E_s/E_t$, which gives
Eq.~\ref{eq:modified-selfish-mining}.
\end{proof}

\fi

\end{document}